\documentclass[%
 reprint,
 amsmath,
 amssymb,
onecolumn,
prb
]{revtex4-1}

\usepackage{graphicx}
\usepackage{dcolumn}
\usepackage{bm}
\usepackage{color}
\usepackage{bbold}

\begin{document}

\preprint{APS/123-QED}

\title{Reply to ``Comment on \\
``Velocity and Speed Correlations in Hamiltonian Flocks"\, ''} 

\author{Mathias Casiulis}
\affiliation{Sorbonne Universit\'{e}, CNRS UMR 7600, Laboratoire de Physique Th\'{e}orique de la Matière Condens\'{e}e, LPTMC, 4 place Jussieu, Couloir 12-13, 5ème \'{e}tage, 75252 Paris Cedex 05, France}
\email{casiulis@lptmc.jussieu.fr}
\author{Marco Tarzia}
\affiliation{Sorbonne Universit\'{e}, CNRS UMR 7600, Laboratoire de Physique Th\'{e}orique de la Matière Condens\'{e}e, LPTMC, 4 place Jussieu, Couloir 12-13, 5ème \'{e}tage, 75252 Paris Cedex 05, France}
\affiliation{Institut Universitaire de France, 1 rue Descartes, 75231 Paris Cedex 05, France}
\author{Leticia F. Cugliandolo}
\affiliation{Sorbonne Universit\'{e}, CNRS UMR 7589, Laboratoire de Physique Th\'{e}orique et Hautes Energies, LPTHE, 4 place Jussieu, Couloir 13-14, 5ème \'{e}tage, 75252 Paris Cedex 05, France}
\affiliation{Institut Universitaire de France, 1 rue Descartes, 75231 Paris Cedex 05, France}
\author{Olivier Dauchot}
\affiliation{UMR Gulliver 7083 CNRS, ESPCI Paris, PSL Research University, 10 rue Vauquelin, 75005 Paris, France}

\date{\today}

\begin{abstract}
In their comment on our work (ArXiv:1912.07056v1), Cavagna \textit{et al.} raise several interesting points on the phenomenology of flocks of birds, and conduct additional data analysis to back up their points.
In particular, they question the existence of rigid body rotations in flocks of birds.
In this reply, we first clarify the notions of rigid body rotations, and of rigidity itself.
Then, we justify why we believe that it is legitimate to wonder about their importance when studying the spatial correlations between speeds in flocks of birds.
\end{abstract}

\maketitle

\section{Introduction}
In their comment on our work~\cite{CavagnaComment}, Cavagna \textit{et al.} raise several interesting points on the phenomenology of flocks of birds, and conduct additional data analysis to back up their points. Before addressing each of these points, let us re-state our observation and claim:
\begin{itemize}
\item First, we show that a 2d Hamiltonian fluid of particles carrying spins coupled to their velocities, present a phase of flocking, while being at thermodynamic equilibrium. This is our main result.
\item Second, although this model is \emph{not} a model of birds, the velocity and speed correlations are strikingly similar to those observed in flocks of birds.  Hence, a natural question arises: in bird flocks, what is the part of the correlations due to, on the one hand, the Goldstone mode related to the polarization of the flock and on the other hand, rotations? As we shall clarify below, it is not obvious which effect is stronger, hence the legitimacy of the question. 
Analyzing a data set available in the literature shows that criticality dominates the velocity correlations, while rotation shapes the speed correlations. We conclude by calling for more systematic data analysis.
\end{itemize}
Cavagna \textit{et al.} comment on these claims, showing that rigid rotations are biologically unattainable and in contrast with all available experimental evidence. Our question is, thus, according to them, not legitimate.\\ 

First, we are glad to see that their comment is the opportunity to present previously unpublished data about the speed correlations as well as about the effect of rotation: this is precisely what our work is calling for.
Second, their comment has enforced us to be more precise in our statements and analysis. This has contributed to the clarity of our message, as reflected by the second version of our paper (arXiv:1911.06042v2), and we thank them for this.
However, for the reasons we shall provide now, our main claim remains unchanged.\\

The points made by Cavagna \textit{et al.} have to do with either of the three following aspects: 
\begin{itemize}
    \item[(i)] the general kinematic properties of sets of points when they undergo a turn or a rotation;
    \vspace{-2mm}
    \item[(ii)] the link between structure and rigidity in sets of particles, in particular, in the case of non-metric interactions;
    \vspace{-2mm}
    \item[(iii)] in the special case of birds, the origin of displacement-displacement correlations.
\end{itemize}

\section{Kinematics of rigid turns and rotations}
We first clarify what we mean exactly by rigid body rotations in our paper, and how it relates to Cavagna \textit{et al.}'s Equal Radius Turns and Parallel Path Turns.
Strictly speaking, rigidity means that the distances between particles remain constant. In practice, there are 
fluctuations and a criterion for rigidity over a chosen time scale $dt$ is that the relative displacements 
satisfy
\begin{equation}
\frac{\Delta | \bm{r}_i - \bm{r}_j| }{|\bm{r}_i-\bm{r}_j|} \ll 1
\qquad\mbox{with}\qquad
\Delta | \bm{r}_i - \bm{r}_j| = | \bm{r}_i - \bm{r}_j|(t+dt) - | \bm{r}_i - \bm{r}_j|(t)
\;,
\label{eq:rigidity}
\end{equation}
where $\Delta | \bm{r}_i - \bm{r}_j|$ is the variation of the length of $\bm{r}_i - \bm{r}_j$ during $dt$.
Let us stress that this definition is purely geometric and does not require nor imply any mechanical property.\\

We can now give the kinematic definitions (independent of their physical origin) of the terms ``parallel path turn", and 
``equal radius turn", that are used in the study of flocks of birds~\cite{Cavagna2018,CavagnaComment}, and that of a ``rigid body rotation".
\begin{itemize}
\item[-]
A $2d$ rigid set of points, $\bm{r}_i$, performs a parallel path turn when it undergoes a 
pure rotation of an angle $d\theta$ around an arbitrary point $P$
with position $\bm{r}_P$:
\begin{equation}
\bm{r}_i(t+d t) - \bm{r}_i(t) =  [(\bm{r}_i(t) -\bm{r}_P) \times \bm{\Omega}(t) ] dt \quad \text{with} \quad \bm{\Omega}(t) dt = d\theta \hat e_z. 
\end{equation}
\item[-]
What is called  a rigid body rotation is a particular case of this motion with $\bm{r}_P = \bm{r}_G$, the position of the
centre of mass $G$. 
\vspace{-2mm}
\item[-]
In an equal radius turn all points turn around different points, $\bm{r}_{P_i}$, with the same radius of curvature $R$
\begin{equation}
\bm{r}_i(t+d t) - \bm{r}_i(t) =  [(\bm{r}_i(t) -\bm{r}_{P_i}) \times \bm{\Omega}(t) ] dt = \bm{R}(t) \times \bm{\Omega}(t)  dt
\;,
\end{equation}
where $\bm{R} = \bm{r}_i(t) -\bm{r}_{P_i}$ is the same vector with length $R$ for all particles.

\end{itemize}
Any rigid transformation can be uniquely decomposed into a translation and 
a rotation around the centre of mass~\cite{Landau1986}. The parallel path turn contains both; 
the rigid body rotation is a pure rotation around $G$; and the equal radius turn is a pure translation.
One should therefore be careful not to confuse \textit{turns}, \textit{i.e.} processes that change the orientation of the velocities, and \textit{rotations}: an equal radius turn, being a pure translation, contains no rotation.

A few additional comments on the list above are in order.
First, this list is \textit{not} an exhaustive list of all the possible rigid transformations: indeed, one can combine pure translations and rotations to create transformations that are not equal radius turns, parallel path turns, or rigid body rotations.
Neither is it an exhaustive list of all the possible turning mechanisms. 
It is purely intended as a lexicon of the few terms used in the present discussion. 
Finally, the list above belongs to rigid transformations only \textit{i.e.} the turns and rotations of non deformable objects.

\begin{figure}
    \centering
    \includegraphics[width=.98\columnwidth]{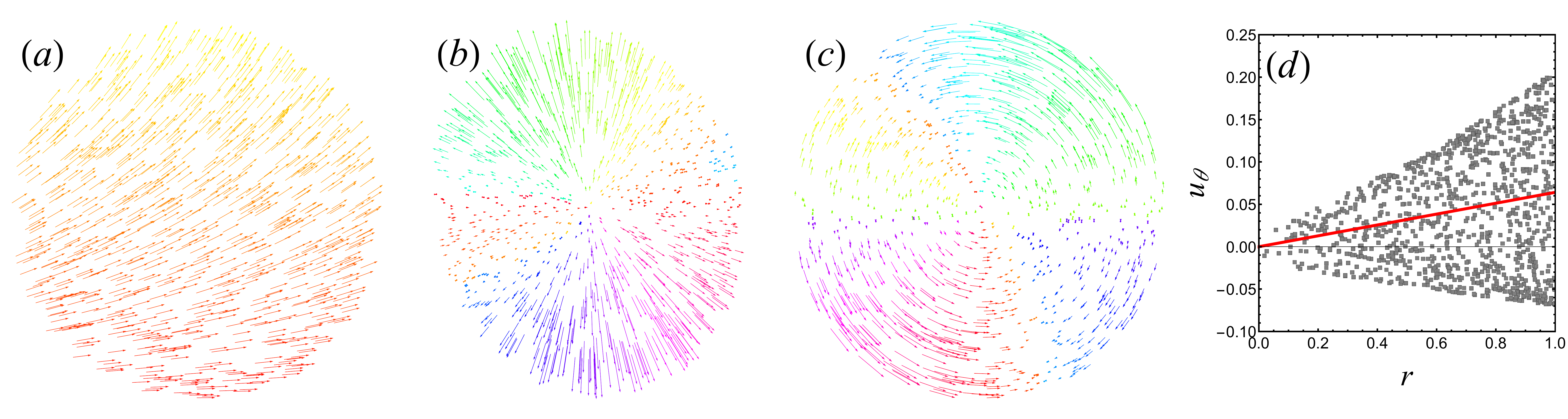}
    \caption{\textbf{Rotation in a spin wave.}
    $(a)$ Displacement field obtained from Eq.~(\ref{eq:SW}), for $1000$ particles uniformly drawn on a unit disk ($L = 2$), with $y_0 = -1$ and $v_0 = 0.1$.
    $(b)$ Radial and $(c)$ orthoradial components of the relative displacement field $\bm{u}^\star = \bm{u} - \bm{u}_G$.
    $(d)$ Orthoradial component of the relative displacement against the distance to the center of the disk (gray), and best linear fit (red).
    }
    \label{fig:SpinWave}
\end{figure}

We can now turn to the field introduced by Cavagna \textit{et al.} in their comment [Eq.~($1$)], as a field which cannot be generated by a solid body rotation; since with fixed speed there is no way a rigid body can turn. 
Although it is clear this field cannot be generated by solid body rotation only; it is just as clear that it contains a solid body rotation component, as we here explicitly check. Take the example provided in their Fig.~$3$, which corresponds to the equations,
\begin{align}
    v_x &= v_0 \cos\left( \frac{(y-y_0) \pi}{3L}\right), \\
    v_y &= v_0 \sin\left( \frac{(y-y_0) \pi}{3L}\right), \label{eq:SW}
\end{align}
with a non-zero phase shift $y_0 \pi / 3 L$ such that the vertical component of the field is not symmetrical around $y = 0$ (otherwise instead of turning the flock splits in two). 
Here we set $v_0 = 0.1$, draw points uniformly on a disk with a unit radius ($L = 2R = 2$), and choose $y_0 = -1$, so that a point at the bottom of the disk has a horizontal velocity.
This field is plotted for $N = 1000$ random points in Fig.~\ref{fig:SpinWave}$(a)$.
From this field, we construct the relative displacement field $\bm{u}^\star = \bm{u} - \bm{u}_G$, where $\bm{u}_G = N^{-1} \sum_i\bm{u}_i$.
Using polar coordinates with an origin at the center of the disk, we decompose the field $\bm{u}^\star$ into its radial (Fig.~\ref{fig:SpinWave}$(b)$) and orthoradial (Fig.~\ref{fig:SpinWave}$(c)$) components.
The latter, $u_{\theta}$, is plotted against the distance to the origin in Fig.~\ref{fig:SpinWave}$(d)$.
The linear drift of this scatterplot is the rotation part of the displacement field.
The best fit (red line in Fig.~\ref{fig:SpinWave}$(d)$) yields a rotation of $\Delta\theta \approx 7.10^{-2} \text{rad}$, or about $4$ degrees.
In other words: this field does contain a rotation around its center of mass, although its vectors all have the same modulus.

\section{Rigidity of flocks}
Cavagna \textit{et al.} quote our sentence ``On short time scales, the flocks do not rearrange: they are solid''. 
On one hand, they agree that on short time scales the flocks do not rearrange~\cite{Cavagna2013}. On the other hand, they disagree with the claim that the flocks are solid. 
We agree that we should not have used the word ``solid"  which refers in the realm of liquid state theory to structural and mechanical properties. We therefore prefer here and in the new version of our paper the word rigid in the  geometrical sense of Eq.~(\ref{eq:rigidity}). 
Given that there is no strong dilatation nor compression of the flock, on short time scales, flocks of birds are rigid in the above sense. Of course, the origin of rigidity is different in flocks of birds and in our model, but this is another issue.

In their comment, Cavagna \textit{et al.} argue that birds cannot display rotations around the center of mass of the flock because, having essentially no structure~\cite{Cavagna2008} and only topological interactions~\cite{Bialek2012}, they cannot be rigid. According to them solid structure always implies a structured g(r), be it crystalline, or glassy. 

While this statement may be true for liquids interacting through standard pairwise potential, the statement is clearly wrong in general: a random set of points on a plane, connected to their nearest neighbors in the Voronoi sense with rigid bars, is a purely topological system, has a flat $g(r)$ and an infinite stiffness. The same will remain true with finite stiffness springs.  Such systems have for instance been studied in the field of amorphous materials~\cite{Amir2013}.
Hence having no structure does not guarantee no rigidity.

\section{Origin of correlations}
We now come to what we believe is the core of the comment on our work by Cavagna et al.
Two questions arise in the context of birds: are there any rotations around the centre of mass in flocks, and if so, what does it mean for the displacement-displacement correlations?

\vspace{0.25cm}
Regarding the presence or absence of rotations, our understanding is the following. 
Clearly there are biological limitations to birds performing sharp turns via parallel path turns. 
Certainly, as Cavagna \textit{et al.} demonstrated, birds perform turns according to the equal radius turn mechanism, \textit{i.e.} translations. However, this does not preclude the presence of some amount of rigid body rotations.

At short enough times, flocks of birds are rigid, as per the definition provided by Eq.~(\ref{eq:rigidity}). 
Therefore, when decomposing the displacement field into a translation, a rigid body  rotation and a, by construction small, deformation, there is no reason to ignore the rigid body rotation however small it might be.
Cavagna et al., believe that insisting on this would be like claiming that any set of experimental points contains some underlying linear law because one can always perform a linear fit. We turn the argument around and claim that ignoring a linear contribution without a firm theoretical reason is even more dangerous and unjustified. As far as we know, in the case of flocks of birds there is no theoretical reason to eliminate this ``linear component''. One solely needs to check that the rotation found is compatible with the physiological abilities of the birds, that is that it remains small.

\vspace{0.25cm}
Regarding the impact that these rotations may have on the shape of the correlations, clear statements are:
\begin{itemize}
\item[-] Equal radius turn are translations and do not produce correlations
\item[-] Any amount of rigid body rotation produces correlations whose only typical scale is the size of the system.
\end{itemize}

It is therefore all a matter of the relative magnitude of the correlations created by rotations and those created by the slow modes of the velocities, seen as $O(n)$ spins. For instance, for a turn such as the one described by Eq.~($2$) of Cavagna \textit{et al.}'s comment, the speed-speed correlation is by construction exactly zero because all speeds are equal. If one were to add an arbitrarily small rotation around the center of mass of this flock, it would be the leading effect.\\

In the case of real flocks instead, it is not obvious which effect is stronger. The rotations are clearly small for the above mentioned physiological reasons.
But given that Goldstone modes themselves lead to correlations with amplitudes that scale like $1 - m^2$, with $m$ the magnetization, and that flocks are typically very polar~\cite{Cavagna2010}, the magnitude of the correlations due to Goldstone modes in flocks is also be expected to be small.\\

Hence the legitimacy of our question.\\

It should be stressed here that, since the correlation functions caused by rotations and by Goldstone modes cancel roughly at the same fraction of the system size, it is not enough to look at the correlation length of Cavagna \textit{et al.}'s Fig.~5 to assess the importance of rotations.
Instead, one should consider the whole correlation functions, and investigate whether their functional forms match either rotations or spin waves more closely, especially at long distances where they differ.
\newpage
\section{Conclusions}
We agree that some statements in the first version of our manuscript were not precise enough, notably regarding the rigidity of flocks, and the relative importance of the correlations caused by rotations and Goldstone modes.

However we do maintain our main claims: (i) given that small solid body rotations are not forbidden by physics nor by physiology, they are most likely present in flocks of birds; (ii) questioning the way they impact the velocity and speed correlation function is legitimate.

In their comment, Cavagna \textit{et al.} made public a variety of functional forms of the correlation functions in different flocks of birds. We are happy to see that the variability is pretty large, which may suggest that the relative contribution of the rotation and the spin waves may vary from one event to another. These newly disclosed data even further call for a systematic decomposition of the displacement fields on translation, rotation and deformation to truly isolate the respective origins of the functional form of the velocity and speed-speed correlations. If, by the end of the analysis one finds that rotations only contribute very little to the correlations, we shall be happy as well.
Our work will have stimulated an important and we believe necessary analysis.

\bibliography{Bibtex-Droplets}

\end{document}